\documentclass[11pt, epsf]{article}

\usepackage{epsfig}
\usepackage{amssymb}
\usepackage{graphicx}
\usepackage{color}
\usepackage{subfigure}
\usepackage{dsfont}

\begin{document}

\baselineskip 6mm
\renewcommand{\thefootnote}{\fnsymbol{footnote}}


\newcommand{\nc}{\newcommand}
\newcommand{\rnc}{\renewcommand}



\newcommand{\tcb}{\textcolor{blue}}
\newcommand{\tcr}{\textcolor{red}}
\newcommand{\tcg}{\textcolor{green}}


\def\be{\begin{equation}}
\def\ee{\end{equation}}
\def\ba{\begin{array}}
\def\ea{\end{array}}
\def\bea{\begin{eqnarray}}
\def\eea{\end{eqnarray}}
\def\nn{\nonumber\\}


\def\ct{\cite}
\def\la{\label}
\def\eq#1{(\ref{#1})}


\def\a{\alpha}
\def\b{\beta}
\def\g{\gamma}
\def\G{\Gamma}
\def\d{\delta}
\def\D{\Delta}
\def\e{\epsilon}
\def\et{\eta}
\def\ph{\phi}
\def\Ph{\Phi}
\def\ps{\psi}
\def\Ps{\Psi}
\def\k{\kappa}
\def\l{\lambda}
\def\L{\Lambda}
\def\m{\mu}
\def\n{\nu}
\def\th{\theta}
\def\Th{\Theta}
\def\r{\rho}
\def\s{\sigma}
\def\S{\Sigma}
\def\ta{\tau}
\def\o{\omega}
\def\O{\Omega}
\def\pr{\prime}


\def\half{\frac{1}{2}}

\def\goto{\rightarrow}

\def\na{\nabla}
\def\grad{\nabla}
\def\curl{\nabla\times}
\def\div{\nabla\cdot}
\def\pa{\partial}
\def\fr{\frac}

\def\bra{\left\langle}
\def\ket{\right\rangle}
\def\lb{\left[}
\def\lc{\left\{}
\def\ls{\left(}
\def\lp{\left.}
\def\rp{\right.}
\def\rb{\right]}
\def\rc{\right\}}
\def\rs{\right)}

\def\vac#1{\mid #1 \rangle}


\def\td#1{\tilde{#1}}
\def\check{ \maltese {\bf Check!}}


\def\Tr{{\rm Tr}\,}
\def\det{{\rm det}}


\def\bc#1{\nnindent {\bf $\bullet$ #1} \\ }
\def\ch {$<Check!>$ }
\def\ss {\vspace{1.5cm}}
\def\text#1{{\rm #1}}
\def\Id{\mathds{1}}

\begin{titlepage}

\hfill\parbox{5cm} { }

\vspace{25mm}

\begin{center}
{\Large \bf Holographic Nucleons in the Nuclear Medium }

\vskip 1. cm
  {Bum-Hoon Lee$^{ab}$\footnote{e-mail : bhl@sogang.ac.kr}} and
    {Chanyong Park$^b$\footnote{e-mail : cyong21@sogang.ac.kr}}

\vskip 0.5cm

{ \it $^a\,$ Department of Physics, Sogang University, Seoul 121-742, Korea}\\
{\it $^b\,$ Center for Quantum Spacetime (CQUeST), Sogang University, Seoul 121-742, Korea}\\

\end{center}

\thispagestyle{empty}

\vskip2cm


\centerline{\bf ABSTRACT} \vskip 4mm

\vspace{1cm}

We investigate the nucleon's rest mass and dispersion relation in the nuclear medium 
which is holographically described by the thermal charged AdS geometry. On this background,
the chiral condensate plays an important role to determine the nucleon's mass
in both the vacuum and the nuclear medium.
It also significantly modifies the nucleon's dispersion relation. 
The nucleon's mass in the high density regime increases with density 
as expected, while in the low density regime it slightly decreases. 
We further study the splitting of the nucleon's masses caused by the isospin interaction 
with the nuclear medium.

\vspace{2cm}


\end{titlepage}

\renewcommand{\thefootnote}{\arabic{footnote}}
\setcounter{footnote}{0}


\section{Introduction}

The AdS/CFT correspondence is a fascinating and useful tool to understand physical 
phenomena. It says that 
the quantum field theory (QFT) in the strong coupling regime 
can be figured out from a classical one-dimensional
higher gravity theory \cite{Maldacena:1997re,Gubser:1998bc,Witten:1998qj,Witten:1998zw}. Many interesting phenomena of the quantum chromodynamics (QCD)
and condensed matter theory happen in the strong coupling regime. 
Therefore, applying the AdS/CFT correspondence to them
may shed light on understanding the nonperturbative aspects of various strongly interacting
QFT \cite{Aharony:1999ti,Klebanov:2000me,Horowitz:2006ct}.

In the QCD and its holographic models, there exists a deconfiment phase transition 
between hadrons and quarks \cite{Erlich:2005qh,Da Rold:2005zs,Karch:2006pv,Sakai:2004cn,Sakai:2005yt}. 
Hadrons are fundamental excitations in the confining phase which usually resides in the strong coupling regime. In a nuclear medium, this confining phase
transits into the deconfining phase above a certain critical temperature and chemical potential 
where hadrons dissolve into quarks \cite{Maldacena:1998im,Rey:1998ik,Park:2009nb,Fadafan:2012qy}. 
In the holographic QCD model, the deconfiment phase 
transition is identified with the Hawking-Page transition of the dual gravity \cite{Herzog:2006ra}.
In this procedure, the deconfining phase maps to a black hole geometry, while
the confining phase corresponds to a non black hole geometry with an appropriate IR modification
\cite{Erlich:2005qh,Da Rold:2005zs,Karch:2006pv}.
In the hard wall model, the thermal AdS (tAdS) space with an IR cutoff corresponds to
the confining phase in the zero density limit. If turning on a nonzero nuclear density,
the dual geometry of the confining phase is generalized to the tAdS with a nonzero electric charge which describes the flavor charge of the dual QCD
\cite{Nakamura:2006xk}-\cite{Colangelo:2012jy}. This geometry was called the thermal
charged AdS (tcAdS) space \cite{Lee:2009bya}. 
This geometry has a singularity at the center. 
However, in the hard wall model we need to introduce an IR cutoff in order to represent
the confinement which prevents all bulk fields from approaching this singularity. 
Therefore, the singularity of tcAdS is not harmful at least in the hard wall model.
On this tcAdS background
the holographic study on the decofinement phase transition have shown 
that the holographic phase diagram is similar to the one expected in particle phenomenology
\cite{Park:2009nb,Lee:2009bya}.

The holographic analysis on the tcAdS space has been further generalized to the case
with two flavor charges by regarding $U(2)$ non-Abelian gauge fields 
\cite{Park:2011zp,Lee:2013oya,Lee:2014gna}. 
In this case, the diagonal time components of these gauge fields are dual
to the number density operators of proton and neutron, so one can interpreted 
a tcAdS geometry as a nuclear medium on the dual QFT.
On this tcAdS space with two flavor symmetries, it was shown that
the deconfinment phase transition and symmetric
energy depend on the number asymmetry between proton and neutron
\cite{Park:2011zp}. 
Furthermore, the meson spectra represented by off-diagonal components of 
gauge fields were also studied \cite{Lee:2013oya,Lee:2014gna}. In general, meson's masses increase with the density 
of the nuclear medium. On the other hand, the isospin interaction reduces the meson mass
when the isospin charge of meson is opposite to the net isospin charge of nuclear medium.
It was also shown that the competition between those two interactions can lead to the pion
condensation \cite{Lee:2013oya,Albrecht:2010eg,Nishihara:2014nva,Nishihara:2014nsa}. 

Similarly, the nucleon's mass spectra have been also investigated in the vacuum corresponding to tAdS \cite{Hong:2006ta,Hong:2007kx,Hong:2007ay,Kim:2009bp,Ahn:2009px,Zhang:2010bn} and further in the isospin medium \cite{Lee:2014xda} 
which includes only the
isospin chemical potential without the nuclear density effect. 
Although the isospin medium provides a good playground to figure out the isospin effect
on nucleon's masses, it is less physical.
In order to understand more realistic nuclear physics phenomena, 
we need to go beyond the isospin medium. In this letter, we will investigate 
nucleon's spectra in the nuclear medium holographically.

The rest of paper is organized as follows. In Sec. 2, we summarize the tcAdS geometry
with two flavor symmetries and explain how five-dimensional fermions living in tcAdS
are reduced to proton and neutron in the dual QFT.
In Sec. 3, we discuss nucleon's rest masses and dispersion relations in the nuclear medium.
We finish this work with some concluding remarks in Sec. 4.


\section{Nucleons in a nuclear medium}

The nuclear matter is composed of two kinds of particles, proton and
neutron, with the baryon and isospin charges. 
In order to describe it holographically, one should take into account a gravity theory 
including at least $U(2)$ flavor symmetry. Here we regard an $U(2)_L \times U(2)_R$  
flavor group to represent parity explicitly. In the dual QFT,
it is related to the chirality of nucleons 
\cite{Hong:2006ta,Henningson:1998cd,Muck:1998rr,Henneaux:1998ch,Contino:2004vy}.
In the hard wall model, the gravity action describing the holographic nuclear medium 
is given by \cite{Lee:2013oya,Lee:2014gna}
\bea
S &=&\int d^{5}x \sqrt{-G}  \left[ 
\frac{1}{2\kappa ^{2}}\left( \mathcal{R}-2\Lambda \right) 
- \frac{1}{4g^{2}}  \ls {F}^{(L)}_{MN} {F}^{(L)MN} + {F}^{(R)}_{MN} {F}^{(R)MN} \rs \rb ,
\label{2}
\eea
where $\Lambda =-6/R^{2}$ is the cosmological constant and the gauge field strengths
for $U(2)_L$ and $U(2)_R$ are given by
\bea
F^{(L)}_{MN} &=& \partial _{M} L_{N} -\partial _{N} L_{M} - i \lb L_M, L_N\rb , \nn
F^{(R)}_{MN} &=& \partial _{M} R_{N} -\partial _{N} R_{M} - i \lb  R_M,  R_N \rb . 
\eea
The nuclear medium can be classified by two quantum numbers, baryon and 
isospin charges \cite{Park:2011zp}. This fact implies that it is sufficient to turn on only diagonal time components 
of the gauge field because they uniquely determine quantum numbers of the nuclear medium.
Their nontrivial values, $V^0_t$ and $V^3_t$, break
the $U(2)_L \times U(2)_R$ flavor group to $U(1)_L^2 \times U(1)_R^2$.
This reduced flavor symmetry group can be further decomposed into 
the symmetric and anti-symmetric combinations, $U(1)_S^2$ and $U(1)_A^2$. 
In this case, the symmetric combination corresponds to a parity even state, 
while the antisymmetric one describes a parity odd state.  
The lowest parity even states are identified with proton and neutron. 
Since the energy of a parity even state is lower than that of a parity 
odd state, it is natural in the low energy regime to consider a nuclear medium
composed of the lowest parity even states \cite{Hong:2006ta,Lee:2014xda}. In the holographic model, it 
can be accomplished by taking $L_M = R_M=- V_M$\footnote{
This convention is different from the one used in  \cite{Lee:2013oya}. However, 
the results in \cite{Lee:2013oya} can be reproduced by defining mesons differently. 
For example, defining charged $\r$-mesons like
\be
\r^{\pm}_m = \fr{1}{\sqrt{2}} \ls v^1_m \mp i v^2_m \rs ,
\ee
reproduces the same meson mass spectrum obtained in \cite{Lee:2013oya}.}.
On this background, the deconfinement phase transition and
the symmetry energy have been studied in \cite{Park:2011zp}. In additions, 
$SU(2)$ meson spectra have been investigated in \cite{Lee:2013oya}.

In the holographic model, the dual operators of $V^0_t$ and $V^3_t$ correspond to
baryon and isospin charge of quark respectively. To see this, let us recall the AdS/CFT correspondence. The dual
operator of the bulk gauge field should have the conformal dimension $3$. One of
candidates is a fermionic current, $\bar{\ps} \g_{\m} \ps$, because a fermionic field
in a $3+1$-dimensional conformal field theory has a conformal dimension $3/2$. 
This fact indicates
that the duals of the bulk gauge fields are not nucleons but quarks. 
However, since fundamental excitations in the confining 
phase are nucleons, one need to reinterpret quark's quantum numbers in terms of nucleon's 
quantities. In the hard wall model representing the confining phase, the previous quark's quantities
can be easily reinterpreted as nucleon's ones by using the conservation of the net quark number.
As a consequence, the resulting tcAdS geometry can be described by \cite{Park:2011zp,Lee:2013oya}
\be		\la{back:geom}
ds^2 = \fr{R^2}{z^2} \ls - f(z) dt^2 + \fr{1}{f(z)} dz^2 + d \vec{x}^2  \rs ,
\ee
with
\bea
f(z) &=& 1 + \fr{3 Q^2 \k^2}{g^2 R^2} z^6 + \fr{D^2 \k^2}{3 g^2 R^2} z^6 , \nn
V^0_t &=& \fr{Q}{\sqrt{2}} \ls 2 z_{IR}^2 - 3 z^2 \rs, \nn
V^3_t &=& \fr{D}{3 \sqrt{2}} \ls 2 z_{IR}^2 - 3 z^2 \rs ,
\eea
where $Q = Q_P + Q_N$ and $D = Q_P - Q_N$ denote the total nucleon number density
and density difference between proton and neutron. Here $Q_p$ and $Q_N$ are
the number of proton and neutron respectively. 

In the confining phase, another important ingredient is the chiral condensate.
In order to see the chiral condensate effect, one should further introduce a complex scalar field $\Ph$ with a negative mass, $- 3/R^2$.
Let us parameterize the complex scalar field as
\be  \la{res:scalarmodulus}
\Ph =  \ph \Id \ e^{i \sqrt{2} \pi} ,
\ee
where $\pi = \pi^i T^i$ with the $SU(2)$ generators, $T^i$.
Then, the modulus $\ph$ can be mapped to the chiral condensate, while $\pi^i$ corresponds 
to the pseudoscalar fluctuations, the so called pions.   
From now on, we set $R=1$ for convenience.
In the dual geometry, \eq{back:geom}, of the nuclear medium the modulus $\ph$
satisfies the following equations of motion \cite{Lee:2013oya}
\be
0 = \fr{1}{\sqrt{-g}} \pa_z \ls \sqrt{-g} g^{zz} \pa_z \ph \rs + 3 \ph ,
\ee
and its solution is given by
\be
\ph (z) = m_q \ z \ _2 F_1 \ls \frac{1}{6} , \half , \frac{2}{3}, - \frac{\ls D^2 + 9 Q^2 \rs   
z^6 }{3 \ N_c} \rs
 + \s \ z^3 \ _2 F_1 \ls \half, \frac{5}{6},\frac{4}{3},  - \frac{\ls D^2 + 9 Q^2 \rs   
z^6 }{3 \ N_c}\rs ,
\ee 
where $m_q$ and $\s$ denotes the current quark mass 
and chiral condensate respectively and $N_c$ is the rank of the gauge group. 
In general, the gravitational backreaction
of the scalar field changes the background geometry. As shown in \cite{Lee:2010dh}, 
it corresponds to $1/N_c$ correction. 
In this letter we ignore the gravitational backreaction of the scalar field, 
as done in the usual hard wall model.

As explained before, the tcAdS geometry is dual to a nuclear medium
composed of the lowest parity even states, proton and neutron. 
In order to describe nucleons in this nuclear medium, 
we should introduce corresponding bulk fields in the tcAdS space. 
Since nucleons are fermions, the corresponding bulk fields should be also 
fermions. Then, bulk fermions in the tcAdS background are governed by
\cite{Hong:2006ta,Kim:2009bp,Ahn:2009px,Zhang:2010bn,Lee:2014xda}
\begin{eqnarray}		\la{act:fermion}
S &=& i \int d^{5}x\sqrt{- G}\left[  \overline{\Psi }^{1} \Gamma ^{M}\nabla_{M}\Psi^{1}
+ \overline{\Psi }^{2} \Gamma ^{M}\nabla_{M}\Psi^{2}  
- m_{1}\overline{\Psi}^{1}\Psi^{1}-m_{2}\overline{\Psi }^{2}\Psi^{2} \rp \nn
&& \qquad \qquad \qquad - \lp g_Y  \ls 
\overline{\Psi}^{1} \Ph \Psi^{2} + \overline{\Psi}^{2} \Ph^{+} \Psi^{1} \rs
\right] ,\label{42}
\end{eqnarray}
where $g_Y$ denotes the Yukawa coupling. 
Since we are interest in nucleons rather than quarks in the confining phase,
the mass of bulk fermions must be $\pm 5/2$ because this value is
related to the conformal dimension of nucleons, $9/2$, in the dual field theory.  
Furthermore, in order to realize the chirality of the $4$-dimensional fermions
from the $5$-dimensional parity under $U(2)_{L} \leftrightarrow U(2)_{R}$, 
we take $m_1 = - m_2= 5/2$. 
Above the covariant derivative $\nabla_{M}$ is defined as
\bea
\na_M \Ps^{1} &=& \ls \pa_M - \fr{i}{4} \o_M - i L_M \rs \Ps^{1} ,  \nn
\na_M \Ps^{2} &=& \ls \pa_M - \fr{i}{4} \o_M - i R_M \rs \Ps^{2} . 
\eea
In this case, $\Psi^{1}$ and $\Psi^{2}$ transform as $\ls \frac{1}{2},0 \rs$ and $\ls 0,\frac{1}{2} \rs$ 
under the flavor group. In general, the Yukawa term couples $\Psi^{1}$ to $\Psi^{2}$ 
and then breaks the chiral symmetry.

The variation of action leads to the following Dirac equations
\begin{eqnarray}		\la{eq:equationferm}
0 &=& \lb e^M_C \G^C \left(\partial _{M} - \frac{i}{4}\omega _{M}^{AB}\Gamma _{AB} +
i  V_M \right) - m_{1} \rb \Psi^{1} 
- g_Y \ph \Ps^{2} , \nn
0 &=& \lb e^M_C \G^C \left(\partial _{M} - \frac{i}{4}\omega _{M}^{AB}\Gamma _{AB} +
i  V_M \right) - m_{2} \rb \Psi^{2} 
- g_Y \ph \Ps^{1} ,
\end{eqnarray}%
where $\Gamma^{AB}=\frac{i}{2}\left[\Gamma^{A},\Gamma^{B}\right]$
and $L_M=R_M= - V_M$ is used. For the well-defined variation, the solutions of the
Dirac equations should satisfy the following boundary boundary condition
\begin{equation}
\lp \delta\overline{\Psi}^{(1,2)}  \Gamma ^{M}\Psi^{(1,2)} \right|^{z_{IR}}_{\e}=0  , \label{48}
\end{equation}
where $z_{IR}$ and $\e$ are the IR and UV cutoff respectively.
Since this Dirac equation is defined on the curved manifold, it is more convenient to introduce 
quantities on the tangent manifold.
The vielbein $e_{M}^{A}$ of the tcAdS space is given by
\be
e_{M}^{A}=  \text{diag} \ls \fr{\sqrt{f(z)}}{z}, \fr{1}{z}, \fr{1}{z}, \fr{1}{z}, \fr{1}{z \sqrt{f(z)}} \rs,
\ee
where $A, B$ and $M,N$ are indices of the tangent and curved manifold respectively.
Then,  non-zero components of spin connection $\omega_{M}^{AB}$ are 
given by 
\be
\omega _{M }^{5A}= \text{diag} \ls \fr{f(z)}{z} - \fr{f(z)'}{2}, 
\fr{\sqrt{f(z)}}{z}, \fr{\sqrt{f(z)}}{z}, \fr{\sqrt{f(z)}}{z}, 0 \rs  . 
\ee
We choose the following gamma matrices on the tangent space
\bea
\G^0 = \ls \begin{array}{cc} 0 &  i \\ i  & 0 \end{array} \rs , \quad 
\G^i = \ls \begin{array}{cc} 0 & - i \s^i \\   i \s^i   & 0 \end{array} \rs , \quad 
\G^4 = \ls \begin{array}{cc} 1 & 0 \\ 0  & - 1 \end{array} \rs  .
\eea
Since $\G^0$ is pure imaginary,  $\bar{\Ps} \Ps$ is not invariant 
under the hermitian conjugation. To make the action invariant under the hermitian
conjugation, $i$ in front of the fermion action was inserted. 
If one further defines the $4$-dimensional gamma matrices
$\g^{\m} = \G^{\m}$ ($\m=0,1,2,3$), 
then the $4$-dimensional chirality operator  is given by $\g^5 = \G^4$.

Now, let us think of the Fourier mode expansion of $5$-dimensional fermions 
\be
\Ps (z,t,\vec{x}) = \sum_{\o_n} \int \fr{ d^3 p}{(2 \pi)^4} \ \Ps (z,\o_n,\vec{p}) \ e^{ - i  \ls \o_n t - \vec{p} \cdot \vec{x} \rs } ,
\ee
where $\Ps$ implies either $\Ps^1$ or $\Ps^2$. 
Since solution of $5$-dimensional Dirac equation usually depends on the parity and isospin charge
it is useful to represent fermions with the parity and isospin quantum numbers.
In terms of $4$-dimensional Weyl spinors, 
$\ps_L$ and $\ps_R$ satisfying $\ps_L = \g^5 \ps_L$ and 
$\ps_R = - \g^5 \ps_R$, the Fourier mode can be further decomposed into
\cite{Lee:2014xda}
\be
\Ps^{1} (z,\o_n,\vec{p}) = \ls \begin{array}{c}
f_L^{1(n,\pm,\pm)} \ \ps^{(n,\pm,\pm)}_L\\
f_R^{1(n,\pm,\pm)}  \ \ps^{(n,\pm,\pm)}_R 
\end{array} \rs  \quad {\rm and} \quad 
\Ps^{2} (z,\o_n,\vec{p})  = \ls \begin{array}{c}
f_L^{2(n,\pm,\pm)} \ \ps^{(n,\pm,\pm)}_L\\
f_R^{2(n,\pm,\pm)}  \ \ps^{(n,\pm,\pm)}_R 
\end{array} \rs ,
\ee
where $n$ denotes the $n$-th resonance
and the first and second sign imply the parity and isospin quantum number respectively.
In these decompositions, the $n$-th mode functions $f$ are given by functions of $z$, $\o_n$ and 
$\vec{p}$.

If one takes the 
normalizable mode functions to be $f^{1}_L$ and $f^{2}_R$, 
the $5$-dimensional parity under the $U(2)_L \times U(2)_R$ flavor group can be associated
with the $4$-dimensional chirality. 
Using the previous Fourier mode decomposition, the $5$-dimensional Dirac equation in \eq{eq:equationferm} is reduced to
\bea
\ls \begin{array}{cc}
{\cal D}_-   \  \mathds{1}& - \fr{g_Y \ph}{z}  \ \mathds{1}\\
- \fr{g_Y \ph}{z} \ \mathds{1} &  {\cal D}_+ \ \mathds{1}
\end{array} \rs  
\ls \begin{array}{c}
f_L^{1(n,\pm,\pm)} \\
f_L^{2(n,\pm,\pm)} 
\end{array} \rs
&=& -
\ls \begin{array}{cc}
\mathbb{E}_+  & 0 \\
0 &  \mathbb{E}_+
\end{array} \rs 
\ls \begin{array}{c}
f_R^{1(n,\pm,\pm)} \\
f_R^{2(n,\pm,\pm)} 
\end{array} \rs , \la{eq:mateq1} \\
\ls \begin{array}{cc}
{\cal D} _+ \ \mathds{1}& \fr{g_Y \ph}{z} \ \mathds{1}\\
 \fr{g_Y \ph}{z} \ \mathds{1}&  {\cal D}_- \ \mathds{1}
\end{array} \rs  
\ls \begin{array}{c}
f_R^{1(n,\pm,\pm)} \\
f_R^{2(n,\pm,\pm)}  
\end{array} \rs
&=& 
\ls \begin{array}{cc}
\mathbb{E}_- & 0 \\
0 &  \mathbb{E}_- 
\end{array} \rs 
\ls \begin{array}{c}
f_L^{1(n,\pm,\pm)}\\
f_L^{2(n,\pm,\pm)} 
\end{array} \rs  , \la{eq:mateq2} 
\eea
where ${\bf 1}$ denotes a $2 \times 2$ identity matrix and
\bea
{\cal D}_{\pm} &=&  \sqrt{f(z)} \lb \pa_z - \fr{2}{z} \ls 1 - \fr{z f'}{8  f(z) } \rs \rb \pm \fr{5}{2z} , \\
\mathbb{E}_{\pm} &=& \fr{1}{\sqrt{f(z)}}  \ls \o_n - V_t \rs \ \mathds{1} \pm  \vec{\s}  \cdot \vec{p} \ . 
\la{eq:matrixeq}
\eea
Above most matrix elements are proportional to the identity matrix
except the last term in \eq{eq:matrixeq}. To solve the Dirac equation,
we first need to determine mode functions as eigenfunctions of $\vec{\s}  \cdot \vec{p}$. By using the rotation symmetry, without loss of generality, we can take the momentum vector to be 
$\vec{p} = \{ 0,0,\pm p \}$.
In this case, mode functions are identified with eigenfunctions with an eigenvalue, $p$ or $-p$.
Now, we take $f^1_L$ and $f^1_R$ to be eingenfuntions with the eigenvalue $p$ 
and $f^2_L$ and $f^2_R$ as eigenfunctions with $-p$. 
Then, \eq{eq:mateq1} and \eq{eq:mateq2} are further simplified to
\bea
\ls \begin{array}{cc}
{\cal D}_-   \  \mathds{1}&  - \fr{g_Y \ph}{z}  \ \mathds{1}\\
- \fr{g_Y \ph}{z} \ \mathds{1} &  {\cal D}_+ \ \mathds{1}
\end{array} \rs  
\ls \begin{array}{c}
f_L^{1(n,\pm,\pm)} \\
f_L^{2(n,\pm,\pm)} 
\end{array} \rs
&=& -
\ls \begin{array}{cc}
E_+ \ \mathds{1} & 0 \\
0 &  E_- \ \mathds{1}
\end{array} \rs 
\ls \begin{array}{c}
f_R^{1(n,\pm,\pm)} \\
f_R^{2(n,\pm,\pm)} 
\end{array} \rs , \la{eq:mateq3} \\
\ls \begin{array}{cc}
{\cal D} _+ \ \mathds{1}& \fr{g_Y \ph}{z} \ \mathds{1}\\
 \fr{g_Y \ph}{z} \ \mathds{1}&  {\cal D}_- \ \mathds{1}
\end{array} \rs  
\ls \begin{array}{c}
f_R^{1(n,\pm,\pm)} \\
f_R^{2(n,\pm,\pm)}  
\end{array} \rs
&=& 
\ls \begin{array}{cc}
E_- \ \mathds{1}& 0 \\
0 &  E_+ \ \mathds{1}
\end{array} \rs 
\ls \begin{array}{c}
f_L^{1(n,\pm,\pm)}\\
f_L^{2(n,\pm,\pm)} 
\end{array} \rs  , \la{eq:mateq4} 
\eea
with
\be
E_{\pm} = \fr{1}{\sqrt{f(z)}}  \ls \o_n - V_t \rs  \pm  p \ .
\ee

In order to identify bulk fermionic components with nucleons of the dual QFT, let us introduce 
symmetric or antisymmetric combinations of mode functions.
Defining the symmetric combination 
\be		\la{con:symmetric}
 f_L^{1(n,+,\pm)} = f_R^{2(n,+,\pm)} \ \text{and} \quad  f_R^{1(n,+,\pm)} = - f_L^{2(n,+,\pm)} ,
\ee
it describes a parity even state. Inserting this symmetric relation into \eq{eq:mateq3}  and \eq{eq:mateq4}, one can easily check that \eq{eq:mateq3}  and \eq{eq:mateq4} are reduced
to the same matrix equation
\bea		\la{eq:protonneutroneq}
\ls \begin{array}{cc}
{\cal D}_-   \  \mathds{1}& \fr{g_Y \ph}{z}  \ \mathds{1}\\
\fr{g_Y \ph}{z} \ \mathds{1} &  {\cal D}_+ \ \mathds{1}
\end{array} \rs  
\ls \begin{array}{c}
f_L^{1(n,+,\pm)} \\
f_R^{1(n,+,\pm)} 
\end{array} \rs
&=&
\ls \begin{array}{cc}
- E_+  & 0 \\
0 & E_-
\end{array} \rs 
\ls \begin{array}{c}
f_R^{1(n,+,\pm)} \\
f_L^{1(n,+,\pm)} 
\end{array} \rs .
\eea
For a parity odd state, we take an antisymmetric combination satisfying
\be   \la{con:antisymmetric}
f_L^{1(n,-,\pm)} = - f_R^{2(n,-,\pm)} \ \text{and} \quad  f_R^{1(n,-,\pm)} = f_L^{2(n,-,\pm)} .
\ee
Then, similar to the parity even case \eq{eq:mateq3}  and \eq{eq:mateq4} reach to 
\bea
\ls \begin{array}{cc}
{\cal D}_-   \  \mathds{1}& - \fr{g_Y \ph}{z}  \ \mathds{1}\\
- \fr{g_Y \ph}{z} \ \mathds{1} &  {\cal D}_+ \ \mathds{1}
\end{array} \rs  
\ls \begin{array}{c}
f_L^{1(n,-,\pm)} \\
f_R^{1(n,-,\pm)} 
\end{array} \rs
&=&
\ls \begin{array}{cc}
- E_+  & 0 \\
0 & E_-
\end{array} \rs 
\ls \begin{array}{c}
f_R^{1(n,-,\pm)} \\
f_L^{1(n,-,\pm)} 
\end{array} \rs .
\eea
The parity even state has lower energy than the parity odd state. In the QCD
proton and neutron correspond to the lowest parity even states. 
From now on, we concentrate on the lowest resonance with $n=1$.
In this case, the mode function with $n=1$, $f^{1(1,+,+)}_{L,R}$ or  
$f^{1(1,+,-)}_{L,R}$ in \eq{eq:protonneutroneq} represents proton or neutron
respectively.
Due to the different isospin charge of nucleons,
the equation in \eq{eq:protonneutroneq} can be further splited into two cases. 
Proton with the isospin charge $1/2$ is governed by
\bea		\la{eq:protoneq}
&& \ls \begin{array}{cc}
 {\cal D}_-   \  \mathds{1}& \fr{g_Y \ph}{z}  \ \mathds{1}\\
\fr{g_Y \ph}{z} \ \mathds{1} &  {\cal D}_+ \ \mathds{1}
\end{array} \rs  
\ls \begin{array}{c}
f_L^{1(1,+,+)} \\
f_R^{1(1,+,+)} 
\end{array} \rs  \nn
&& \qquad =
\ls \begin{array}{cc}
-  \lc \fr{1}{\sqrt{f(z)}}  \ls \o - \fr{V^0_t  + V_t^3}{2} \rs  +  p   \rc   & 0 \\
0 & \fr{1}{\sqrt{f(z)}}  \ls \o - \fr{V^0_t  + V_t^3}{2} \rs  - p  
\end{array} \rs 
\ls \begin{array}{c}
f_R^{1(1,+,+)} \\
f_L^{1(1,+,+)} 
\end{array} \rs ,
\eea
while for neutron  with the isospin charge of $-1/2$  \eq{eq:protonneutroneq} yields
\bea		\la{eq:neutroneq}
&& \ls \begin{array}{cc}
 {\cal D}_-   \  \mathds{1}&  \fr{g_Y \ph}{z}  \ \mathds{1}\\
 \fr{g_Y \ph}{z} \ \mathds{1} &  {\cal D}_+ \ \mathds{1}
\end{array} \rs  
\ls \begin{array}{c}
f_L^{1(1,+,-)} \\
f_R^{1(1,+,-)} 
\end{array} \rs  \nn
&& \qquad  =
\ls \begin{array}{cc}
-  \lc \fr{1}{\sqrt{f(z)}}  \ls \o - \fr{V^0_t  - V_t^3}{2} \rs  +  p   \rc   & 0 \\
0 & \fr{1}{\sqrt{f(z)}}  \ls \o - \fr{V^0_t - V_t^3}{2} \rs  - p  
\end{array} \rs 
\ls \begin{array}{c}
f_R^{1(1,+,-)} \\
f_L^{1(1,+,-)} 
\end{array} \rs ,
\eea
where we use $\o=\o_1$ for simplicity.
Taking $V^0_t =0$, $V^3_t = \text{const}$ and $f(z)=1$, above equations 
reduces to those for nucleons in the isospin medium \cite{Lee:2014xda}.
In the nuclear medium, unlike  the isospin medium, the energy and mass of nucleons
crucially depends on the medium due to the nontrivial radial coordinate dependence of 
the metric and background gauge fields.

\section{Nucleon spectrum in the nuclear medium}

At given $Q$, $D$, $m_q$ and $\s$, the energy and momentum of nucleons can be determined
by solving \eq{eq:protoneq} or \eq{eq:neutroneq} together with appropriate two boundary 
conditions. For the well-defined variation of the fermionic action, \eq{48} should vanish. 
To do so, we impose the following two boundary conditions
\be
f_L^{1(n,\pm,\pm)}  (0) = 0 \quad {\rm and} \quad f_R^{1(n,\pm,\pm)}  (z_{IR}) = 0  ,
\ee
which was also used in studying the nucleon mass in the isospin medium
\cite{Hong:2006ta,Lee:2014xda}. 
In general, solving the Dirac equation with above boundary conditions 
gives rise to a relation between parameters. Inversely, this fact implies that 
there exists a solution only in the case satisfying this parameter relation.
Furthermore, since the range of $z$ is restricted to $0 \le z \le z_{IR}$,
the solution of the Dirac equation has a discrete eigenvalues. this is why we take
a discrete energy, $\o_n$, rather than a continuous one.
As a consequence, the parameter relation obtained by solving the Dirac equation
is nothing but the dispersion relation of nucleons because it expresses a discrete energy
as a function of the other quantities.
In this case, the rest mass of nucleon appears in the $p=0$ limit. In general,
the dispersion relation crucially depends on properties of the nuclear medium, 
$Q$ and $D$.
In this section, we will investigate how the dispersion relation of nucleons changes
in the nuclear medium.

\subsection{Dispersion relation in the vacuum}

Before studying the nucleon's spectra in the medium, 
let's first consider the vacuum with $Q=D=0$ in order to get more intuitions.
In this case, the dual geometry 
is given by a tAdS space and proton and neutron become degenerate. 
If we further set $m_q=\s=0$, the lowest nucleons with the energy
are governed by
\bea 		\la{eq:reladis}
\bar{{\cal D}}_+ \bar{{\cal D}}_- f^1_L = - (\o^2 - p^2) f^1_L , \nn
\bar{{\cal D}}_- \bar{{\cal D}}_+ f^1_R= - (\o^2 - p^2) f^1_R  ,
\eea
where $\bar{{\cal D}}_+ = \pa_z + \fr{1}{2 z}$ and $\bar{{\cal D}}_- =\pa_z - \fr{9}{2 z}$.
Solutions of these equations depend only the value of $\o^2 - p^2$. 
Suppose that there exists a solution at a given value of $\sqrt{\o^2 - p^2}$.
Denoting this value by $ m_0 = \sqrt{\o^2 - p^2}$,
$m_0$ determines the nucleon's dispersion relation uniquely. In this case,
nucleons follow the relativistic dispersion relation
\be
\o^2 = m_0^2 + p^2 .
\ee
Since $\o$ reduces to $m_0$ at $p=0$, $m_0$ can be identified with the nucleon's
rest mass. This relativistic dispersion relation is expected from 
the asymptotic symmetry of the tAdS geometry.
Since the boundary space of tAdS is invariant under boundary Poincare symmetry, nucleons
defined on this boundary should satisfy the relativistic dispersion relation. To check this, 
we numerically solve \eq{eq:reladis} for $m_q=\s=Q=D=0$.
Numerical results for the nucleon's energy are plotted in Fig. 1(a). 
The resulting curve is well fitted by the
following dispersion relation
\be
\o = \sqrt{2.0589^2 + p^2} .
\ee
This result shows the exact relativistic
dispersion relation and indicates that the nucleon's rest mass in the vacuum without a
chiral condensate is given by $m_0=2.0589$GeV, which is very larger than the real
nucleon's mass. However, as will be shown, the chiral condensation can reduce this large
mass to the real mass.

\begin{figure}
\begin{center}
\vspace{0cm}
\hspace{-0.5cm}
\subfigure[$\ph = 0$]{ \includegraphics[angle=0,width=0.45\textwidth]{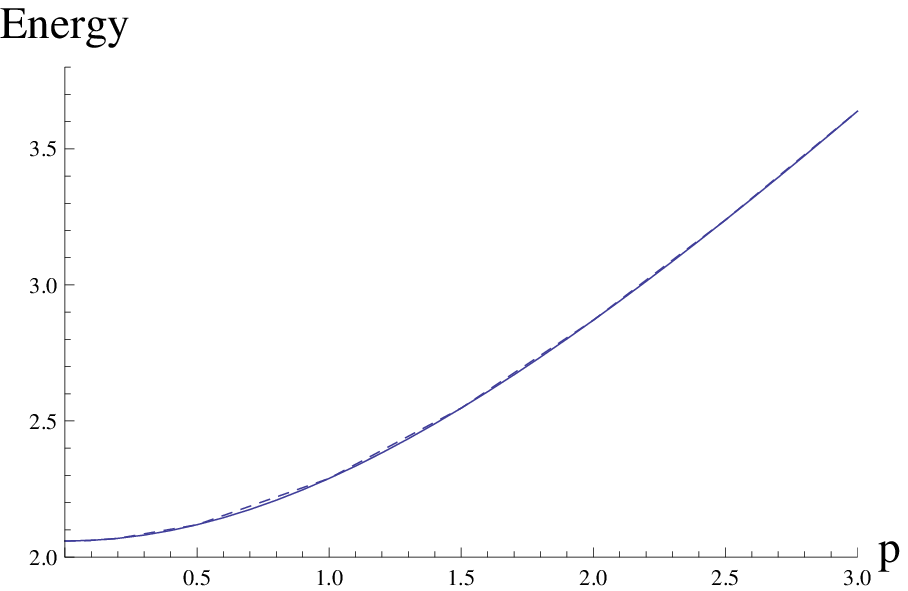}}
\hspace{0cm}
\subfigure[$\ph \ne 0$]{ \includegraphics[angle=0,width=0.45\textwidth]{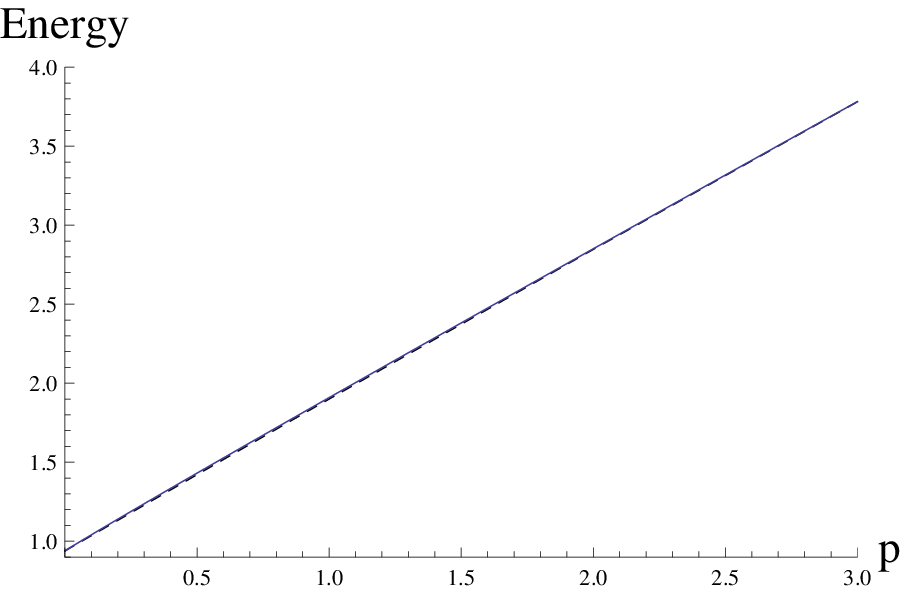}}
\vspace{-0cm}
\caption{\small The nucleon's mass in the vacuum (a) for $m_q=\s=0$ and  (b) with 
$m_q=2.38$MeV, $\s=(304 {\rm MeV})^3$ and $g_Y=4.699$ which
reproduce the correct nucleon's mass in the vacuum. }
\label{number}
\end{center}
\end{figure}

Now, let us consider the effects of the current quark mass and the chiral condensate.
The current quark mass breaks the chiral symmetry explicitly, while the chiral condensation
breaks it spontaneously. To distinguish those two effects, let us first
turn on the current quark mass without the chiral condensate, $m_q \ne 0$ and $\s=0$.
Then, the previous relativistic dispersion relation is slightly modified into
\be
\o^2 = m_0^2 + \ls p+ g_Y m_q \rs^2.
\ee
In the $p=0$ limit, the current quark mass slightly changes the nucleon's mass into
$m^2 = m_0^2 +g_Y^2 m_q^2$. In the large momentum limit where
$p \gg  g_Y m_q$, however, the modified dispersion relation still remains as the relativistic 
one, $\o \sim p$. It was also shown that
the meson's dispersion relation, regardless of $m_q$ and $\s$, 
follows the similar relativistic form in the high momentum region \cite{Lee:2014gna}. 

If the chiral condensate is also turned on, 
the nucleon's dispersion relation is totally changed even in the vacuum. 
In Fig. 1(b), the nucleon's dispersion relation in the vacuum with a chiral condensate is depicted.
Intriguingly,  the obtained dispersion relation is well fitted by
\be
\o = 0.9390 + 0.97 \  p^{0.979} .
\ee
As shown in this result, the chiral condensate dramatically reduces the nucleon's mass to the real 
mass of nucleons,
from $2.0589$GeV to $0.9390$GeV.
Another interesting point is that the chiral condensate modifies the momentum dependence 
in the small momentum limit, from $p^2$ to $p^{0.979}$. 
This is the story in vacuum with $Q=D=0$, 
where there is no distinct between proton and neutron
due to the absence of the isospin interaction.

\subsection{Nucleon's rest mass in the nuclear medium}

Now, let us consider nucleons in the nuclear medium.
As shown in the previous section, the chiral condensate plays a crucial role in determining
the nucleon spectrum so that from now on we focus on the case with
$m_q = 2.38$MeV, , $\s=(304 {\rm MeV})^3$ and $g_Y=4.699$.
The rest masses of nucleons are determined from the energy in the zero momentum limit. 
In order see the the nuclear density effect on the nucleon mass, we first turn off
the isospin interaction by taking $D=0$ but $Q \ne 0$. In this case, 
because of absence of the isospin interaction with the nuclear medium,
proton and neutron are still degenerate. In Fig. 2(a), we plot the nucleon's rest mass depending
on the nuclear density. In the low density regime below a certain critical density, 
the nucleon mass slowly decreases with increasing nuclear density, whereas it rapidly increases 
in the high density region.

\begin{figure}
\begin{center}
\vspace{0cm}
\hspace{-0.5cm}
\subfigure[$D=0$]{ \includegraphics[angle=0,width=0.45\textwidth]{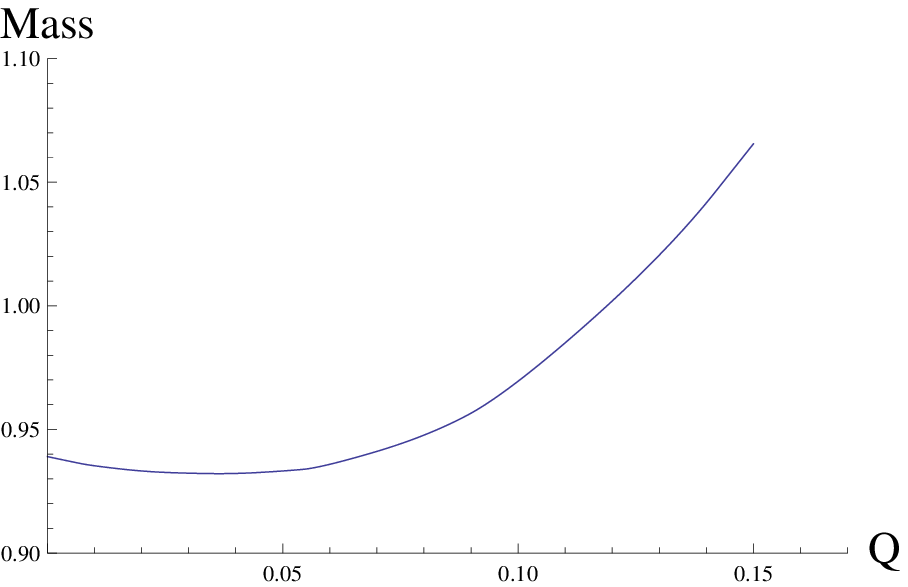}}
\hspace{0cm}
\subfigure[$D=Q/2$]{ \includegraphics[angle=0,width=0.45\textwidth]{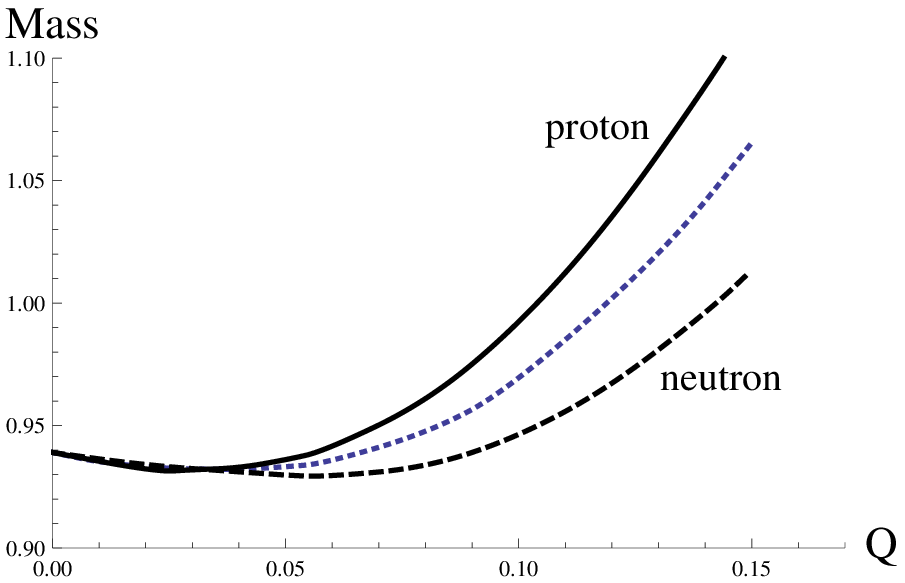}}
\vspace{-0cm}
\caption{\small The nucleon's mass spectrum in the nuclear medium. (a) For $D=0$, proton and
neutron are degenerate. (b) For $D = Q/2$, the masses of proton and neutron are splited due
to the isospin interaction, where the dotted line denotes the nucleon mass for $D=0$.}
\label{number}
\end{center}
\end{figure}

In general cases with a nontrivial isospin interaction with the nuclear medium, 
masses of proton and nucleon are
splitted and become non-degenerate like the meson's spectra. For $D=Q/2$ which
describes the nuclear medium composed of $75\%$ protons and $25\%$ neutrons, 
the density dependence of nucleon's mass is depicted in Fig. 2(b). 
In the high density regime, the proton mass increases more rapidly than the neutron mass.
In the high density regime the isospin interaction 
prefers creation of neutron rather than proton
because in the nuclear medium with the positive net isospin charge
more energy cost is required to create proton as expected.
In the low density region, 
the proton mass decreases slightly faster than the neutron mass unlike the high density case.
Another intriguing result is that nucleon has the lowest mass not at the zero density 
but at a certain critical density. 

\subsection{Dispersion relations}

In the nonzero momentum limit, as mentioned before, 
the energy of the nucleons should be related to their momentum
in order to satisfy the dispersion relation in the nuclear medium.
This dispersion relation usually includes information for the interaction between nucleons
and the background nuclear matter. In this section, after solving \eq{eq:protoneq} 
and \eq{eq:neutroneq} with a nonzero momentum, we investigate effects of the nuclear
density and the isospin interaction on the nucleon's dispersion relations. 
To do so, it should be noted that, when we describe the nuclear medium
in the confining phase, there exists an upper bound in $Q$. For example, the deconfinement 
phase transition occurs at the critical value $Q_c$,
$Q_c = 0.1679$ for $\a=1/2$ and $Q_c = 0.1615$ for $\a=1$ \cite{Lee:2013oya}. 
Therefore, we should restrict the range of $Q$ to $0 \le Q  < Q_c$ for representing
the confining phase. First, we pick up $Q=0.1$ and $\a=0$ to see only the nuclear density effect.
In this case, since $D=0$, there is no distinction between proton and neutron. 
The effect of the nuclear density on the nucleon's dispersion relation is plotted in Fig. 3(a),
where the background nuclear density uplifts the nucleon's energy. In the small momentum
limit, the dispersion relation is fitted by an almost linear curve
\be
\o = 0.9695 + 1.650 \ p^{0.999}  ,
\ee
where $0.9695$GeV is the rest mass of nucleon at $Q=0.1$ and $D=0$. 
Fig. 3(b) shows the splitting of the nucleon's energy when we turn on $\a = 1/2$ with $Q=0.1$. 
Similar to the meson case, the isospin interaction with the nuclear medium
breaks the degeneracy of nucleons. Comparing it with Fig. 3(a), the isospin
interaction increases the proton energy slightly, whereas the neutron's energy decreases.

\begin{figure}
\begin{center}
\vspace{0cm}
\hspace{-0.5cm}
\subfigure[$Q=0.1$ and $\a=0$]{ \includegraphics[angle=0,width=0.45\textwidth]{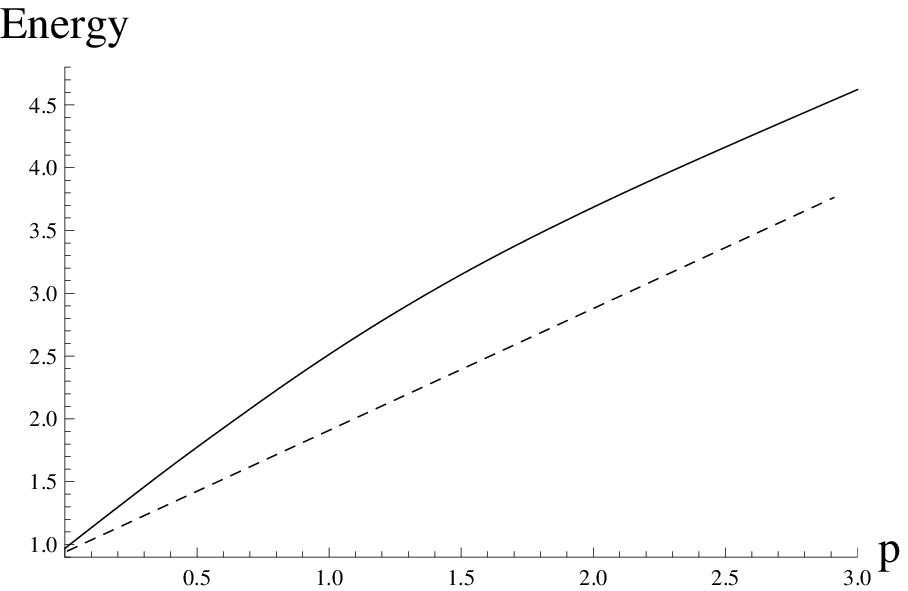}}
\hspace{0cm}
\subfigure[$Q=0.1$ and $\a=1/2$]{ \includegraphics[angle=0,width=0.45\textwidth]{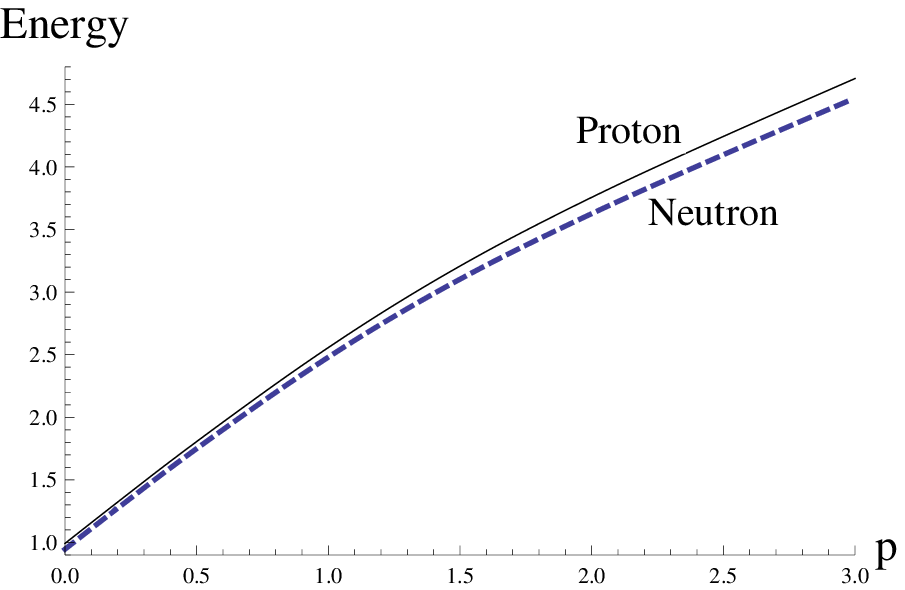}}
\vspace{-0cm}
\caption{\small The nucleon's dispersion relation in the nuclear medium. (a) The dashed
and solid line indicate the dispersion relations in the vacuum and in the nuclear medium 
with the chiral condensate respectively. (b) The isospin interaction splits the degeneracy
of nucleons. The energy of proton (neutron) slightly increases (decreases).}
\label{number}
\end{center}
\end{figure}


\section{Discussion}

In this letter, we have studied the nucleon's rest mass and dispersion relation in the nuclear
medium by using the AdS/CFT correspondence. To describe the nuclear medium with
the flavor symmetry of the dual QFT, we introduced bulk gauge fields of $U(1)_L^2 \times U(1)_R^2 \subset U(2)_L \times U(2)_R$.
These bulk gauge fields are dual to the quark number and isospin operator. In the confining phase,
since nucleons rather than quarks are fundamental, we rewrote 
bulk gauge fields in terms of nucleon quantities by using the conservation of the net quark number,
which uniquely determine the component ratio of proton and neutron in the nuclear medium.

On this background, we turned on the $5$-dimensional fermionic fluctuations with mass,
$\pm 5/2$, and reinterpreted them as $4$-dimensional nucleons, which satisfy the 
$5$-dimensional Dirac equation together with appropriate two boundary conditions. 
By solving the
Dirac equation numerically, we have investigated rest masses and dispersion relations of
the lowest parity even states, proton and neutron. We found that the chiral condensate
is crucial to explain the nucleon's rest mass because it dramatically changes the dispersion relation 
of nucleons unlike meson's spectra studied in \cite{Lee:2014gna}. We also showed 
that in the high nuclear density regime, as expected, nucleon's rest mass increases with 
nuclear density, while in the low density regime it decreases unexpectedly.
It would be interesting to figure out why such an unexpected nucleon's mass spectrum
occurs in the low density regime.

We also showed that the number asymmetry between proton and neutron causes
the mass and energy splitting between proton and neutron, which are similar 
to the meson mass splitting in the nuclear medium \cite{Lee:2013oya} 
and to the nucleon mass splitting in the isospin medium \cite{Lee:2014xda}. 
In the nuclear medium with the relative abundances of protons,
the isospin interactions makes the proton's rest mass and energy at a given momentum
larger than those of neutron. These results would be helpful to understand nucleons
in the nuclear medium quantitatively and qualitatively because there is no QFT tools applicable in the strong coupling regime.

\vspace{1cm}

{\bf Acknowledgement} 

This work was supported by the National Research Foundation of Korea (NRF) grant funded by the Korea government (MSIP) (2014R1A2A1A01002306).
C.Park was also supported by Basic Science Research Program through the National Research Foundation of Korea funded by the Ministry of Education (NRF-2013R1A1A2A10057490).

\vspace{1cm}



\begin{thebibliography}{99}

\bibitem{Maldacena:1997re} 
  J.~M.~Maldacena,
  Adv.\ Theor.\ Math.\ Phys.\  {\bf 2}, 231 (1998)
  [hep-th/9711200].
  
\bibitem{Gubser:1998bc} 
  S.~S.~Gubser, I.~R.~Klebanov and A.~M.~Polyakov,
  Phys.\ Lett.\ B {\bf 428}, 105 (1998)
  [hep-th/9802109].
  
\bibitem{Witten:1998qj} 
  E.~Witten,
  Adv.\ Theor.\ Math.\ Phys.\  {\bf 2}, 253 (1998)
  [hep-th/9802150].
  
\bibitem{Witten:1998zw} 
  E.~Witten,
  Adv.\ Theor.\ Math.\ Phys.\  {\bf 2}, 505 (1998)
  [hep-th/9803131].
  
\bibitem{Aharony:1999ti} 
  O.~Aharony, S.~S.~Gubser, J.~M.~Maldacena, H.~Ooguri and Y.~Oz,
  Phys.\ Rept.\  {\bf 323}, 183 (2000)
  [hep-th/9905111].

\bibitem{Klebanov:2000me} 
  I.~R.~Klebanov,
  hep-th/0009139.

\bibitem{Horowitz:2006ct} 
  G.~T.~Horowitz and J.~Polchinski,
  In *Oriti, D. (ed.): Approaches to quantum gravity* 169-186
  [gr-qc/0602037].
  
\bibitem{Erlich:2005qh} 
  J.~Erlich, E.~Katz, D.~T.~Son and M.~A.~Stephanov,
  Phys.\ Rev.\ Lett.\  {\bf 95}, 261602 (2005)
  [hep-ph/0501128].
   
\bibitem{Da Rold:2005zs}
  L.~Da Rold and A.~Pomarol,
  Nucl.\ Phys.\ B {\bf 721}, 79 (2005)
  [hep-ph/0501218].
         
\bibitem{Karch:2006pv} 
  A.~Karch, E.~Katz, D.~T.~Son and M.~A.~Stephanov,
  Phys.\ Rev.\ D {\bf 74}, 015005 (2006)
  [hep-ph/0602229].
    
\bibitem{Sakai:2004cn} 
  T.~Sakai and S.~Sugimoto,
  Prog.\ Theor.\ Phys.\  {\bf 113}, 843 (2005)
  [hep-th/0412141].
  
\bibitem{Sakai:2005yt} 
  T.~Sakai and S.~Sugimoto,
  Prog.\ Theor.\ Phys.\  {\bf 114}, 1083 (2005)
  [hep-th/0507073].
  
\bibitem{Maldacena:1998im} 
  J.~M.~Maldacena,
  Phys.\ Rev.\ Lett.\  {\bf 80}, 4859 (1998)
  [hep-th/9803002].
  
\bibitem{Rey:1998ik} 
  S.~J.~Rey and J.~T.~Yee,
  Eur.\ Phys.\ J.\ C {\bf 22}, 379 (2001)
  [hep-th/9803001].
  
\bibitem{Park:2009nb} 
  C.~Park,
  Phys.\ Rev.\ D {\bf 81}, 045009 (2010)
  [arXiv:0907.0064 [hep-ph]].
  
\bibitem{Herzog:2006ra} 
  C.~P.~Herzog,
  Phys.\ Rev.\ Lett.\  {\bf 98}, 091601 (2007)
  [hep-th/0608151].

\bibitem{Fadafan:2012qy} 
  K.~B.~Fadafan and E.~Azimfard,
  Nucl.\ Phys.\ B {\bf 863}, 347 (2012)
  [arXiv:1203.3942 [hep-th]].
    
\bibitem{Nakamura:2006xk} 
  S.~Nakamura, Y.~Seo, S.~J.~Sin and K.~P.~Yogendran,
  J.\ Korean Phys.\ Soc.\  {\bf 52}, 1734 (2008)
  [hep-th/0611021].
  
\bibitem{Nakamura:2007nx} 
  S.~Nakamura, Y.~Seo, S.~J.~Sin and K.~P.~Yogendran,
  Prog.\ Theor.\ Phys.\  {\bf 120}, 51 (2008)
  [arXiv:0708.2818 [hep-th]].
  
\bibitem{Domokos:2007kt} 
  S.~K.~Domokos and J.~A.~Harvey,
  Phys.\ Rev.\ Lett.\  {\bf 99}, 141602 (2007)
  [arXiv:0704.1604 [hep-ph]].
  
\bibitem{Lee:2009bya} 
  B.~-H.~Lee, C.~Park and S.~J.~Sin,
  JHEP {\bf 0907}, 087 (2009)
  [arXiv:0905.2800 [hep-th]].
  
\bibitem{Jo:2009xr} 
  K.~Jo, B.~-H.~Lee, C.~Park and S.~J.~Sin,
  JHEP {\bf 1006}, 022 (2010)
  [arXiv:0909.3914 [hep-ph]].
  
\bibitem{Kim:2010dp} 
  Y.~Kim, Y.~Seo, I.~J.~Shin and S.~J.~Sin,
  JHEP {\bf 1106}, 011 (2011)
  [arXiv:1011.0868 [hep-ph]].
  
\bibitem{Colangelo:2010pe} 
  P.~Colangelo, F.~Giannuzzi and S.~Nicotri,
  Phys.\ Rev.\ D {\bf 83}, 035015 (2011)
  [arXiv:1008.3116 [hep-ph]].
  
\bibitem{Cai:2012xh} 
  R.~G.~Cai, S.~He and D.~Li,
  JHEP {\bf 1203}, 033 (2012)
  [arXiv:1201.0820 [hep-th]].
  
\bibitem{Colangelo:2012jy} 
  P.~Colangelo, F.~Giannuzzi and S.~Nicotri,
  JHEP {\bf 1205}, 076 (2012)
  [arXiv:1201.1564 [hep-ph]].
  
\bibitem{Park:2011zp} 
  C.~Park,
  Phys.\ Lett.\ B {\bf 708}, 324 (2012)
  [arXiv:1112.0386 [hep-th]].
  
\bibitem{Lee:2013oya} 
  B.~-H.~Lee, S.~Mamedov, S.~Nam and C.~Park,
  JHEP {\bf 1308}, 045 (2013)
  [arXiv:1305.7281 [hep-th]].
  
\bibitem{Lee:2014gna} 
  B.~-H.~Lee, C.~Park and S.~Nam,
  arXiv:1412.3097 [hep-ph].
  
\bibitem{Albrecht:2010eg} 
  D.~Albrecht and J.~Erlich,
  Phys.\ Rev.\ D {\bf 82}, 095002 (2010)
  [arXiv:1007.3431 [hep-ph]].
  
\bibitem{Nishihara:2014nva} 
  H.~Nishihara and M.~Harada,
  Phys.\ Rev.\ D {\bf 89}, no. 7, 076001 (2014)
  [arXiv:1401.2928 [hep-ph]].
  
\bibitem{Nishihara:2014nsa} 
  H.~Nishihara and M.~Harada,
  Phys.\ Rev.\ D {\bf 90}, no. 11, 115027 (2014)
  [arXiv:1407.7344 [hep-ph]].
  
\bibitem{Hong:2006ta} 
  D.~K.~Hong, T.~Inami and H.~-U.~Yee,
  Phys.\ Lett.\ B {\bf 646}, 165 (2007)
  [hep-ph/0609270].
  
\bibitem{Hong:2007kx} 
  D.~K.~Hong, M.~Rho, H.~U.~Yee and P.~Yi,
  Phys.\ Rev.\ D {\bf 76}, 061901 (2007)
  [hep-th/0701276 [HEP-TH]].
  
\bibitem{Hong:2007ay} 
  D.~K.~Hong, M.~Rho, H.~U.~Yee and P.~Yi,
  JHEP {\bf 0709}, 063 (2007)
  [arXiv:0705.2632 [hep-th]].
  
\bibitem{Kim:2009bp} 
  H.~C.~Kim, Y.~Kim and U.~Yakhshiev,
  JHEP {\bf 0911}, 034 (2009)
  [arXiv:0908.3406 [hep-ph]].
  
\bibitem{Ahn:2009px} 
  H.~C.~Ahn, D.~K.~Hong, C.~Park and S.~Siwach,
  Phys.\ Rev.\ D {\bf 80}, 054001 (2009)
  [arXiv:0904.3731 [hep-ph]].
  
\bibitem{Zhang:2010bn} 
  P.~Zhang,
  Phys.\ Rev.\ D {\bf 81}, 114029 (2010)
  [arXiv:1002.4352 [hep-ph]].
  
\bibitem{Lee:2014xda} 
  B.~-H.~Lee, S.~Mamedov and C.~Park,
  Int.\ J.\ Mod.\ Phys.\ A {\bf 29}, no. 29, 1450170 (2014)
  [arXiv:1402.6061 [hep-th]].
  
\bibitem{Henningson:1998cd} 
  M.~Henningson and K.~Sfetsos,
  Phys.\ Lett.\ B {\bf 431}, 63 (1998)
  [hep-th/9803251].
  
\bibitem{Muck:1998rr} 
  W.~Mueck and K.~S.~Viswanathan,
  Phys.\ Rev.\ D {\bf 58}, 041901 (1998)
  [hep-th/9804035].
  
\bibitem{Henneaux:1998ch} 
  M.~Henneaux,
  In *Tbilisi 1998, Mathematical methods in modern theoretical physics* 161-170
  [hep-th/9902137].
  
\bibitem{Contino:2004vy} 
  R.~Contino and A.~Pomarol,
  JHEP {\bf 0411}, 058 (2004)
  [hep-th/0406257].

\bibitem{Lee:2010dh} 
  B.~-H.~Lee, C.~Park and S.~Shin,
  JHEP {\bf 1012}, 071 (2010)
  [arXiv:1010.1109 [hep-th]].


    




  
  
\end{thebibliography}
\end{document}